\documentclass[twocolumn,showpacs,preprintnumbers,amsmath,amssymb,superscriptaddress,balancelastpage,prb]{revtex4}
\include{options}
\usepackage {amsmath,amssymb,epsfig,color}
\usepackage{natbib}
\usepackage {graphicx}

\begin{document}

\title{Phonon-assisted insulator-metal transitions in correlated systems driven by doping}

\author{E.~I.~Shneyder}
\email{eshneyder@gmail.com}
\affiliation{Kirensky Institute of Physics, Federal Research Center KSC SB RAS, 660036 Krasnoyarsk, Russia}
\author{M.~V.~Zotova}
\affiliation{Kirensky Institute of Physics, Federal Research Center KSC SB RAS, 660036 Krasnoyarsk, Russia}
\affiliation{Siberian Federal University, 660041 Krasnoyarsk, Russia}
\author{S.~V.~Nikolaev}
\affiliation{Kirensky Institute of Physics, Federal Research Center KSC SB RAS, 660036 Krasnoyarsk, Russia}
\affiliation{Siberian Federal University, 660041 Krasnoyarsk, Russia}
\author{S.~G.~Ovchinnikov}
\affiliation{Kirensky Institute of Physics, Federal Research Center KSC SB RAS, 660036 Krasnoyarsk, Russia}
\affiliation{Siberian Federal University, 660041 Krasnoyarsk, Russia}
\date{\today}

\begin {abstract}
We consider how electron-phonon interaction influences the insulator-metal transitions driven by doping in the strongly correlated system. Using the polaronic version of the generalized tight-binding method, we investigate a multiband two-dimensional model taking into account both Holstein and Su–Schrieffer–Heeger types of electron-lattice contributions. For adiabatic ratio  $t \gg \omega_0 $, different types of band structure evolution are observed in a wide electron-phonon parameter range. We demonstrate the relationship between transition features and such properties of the system as the polaron and bipolaron crossovers, pseudogap behavior of various origin, orbital selectivity, and the redistribution of the spectral weight due to the electron-phonon interaction.
\end {abstract}

\maketitle

\section{Introduction \label{intro}}
The mutual influence between electron-electron and electron-phonon interactions is one of the intriguing problems of condensed matter theory. It acquires particular importance in the context of the insulator-metal or insulator-semiconductor transitions in the compounds of transition metals~\cite{10.1038/s41586-019-1824-9,10.1038/s42005-020-0330-6,Baldini6409,PARIJA20201166}. Achieving conduction switching in correlated systems with minimal energy dissipation can pave the way towards novel memory elements, low-power neuromorphic computing, or other highly energy-efficient applications \cite{ANDREWS2019711,10.1038/s41467-020-16752-1,10.1038/ncomms8812}. Indeed, many oxides, sulfides, nickelates, and other systems of d-metal exhibit a sharp transition from nonmetal to a metal ground state induced by relatively small alternation in temperature, pressure, strain, composition, or chemical doping~\cite{RevModPhys.40.714,RevModPhys.70.1039,ANDREWS2019711,10.1038/s42005-020-0330-6}. In electronic devices, it can be driven by optical pumping \cite{10.1038/srep25538} or applying an electrical current or voltage~\cite{10.1038/40363}. The transition results in enormous variations of conductivity and related physical properties. However, the dependence of the transformation characteristics on the atomistic and electronic structures of materials remains the outstanding fundamental question. Despite considerable attention to the problem, the underlying mechanisms are the subject of active discussions. These can be structural distortions, electronic instabilities, or their cooperative or competing contributions~\cite{ANDREWS2019711}.

Transition metal compounds belong to a vast class of correlated materials. Due to significant Coulomb interaction, a number of them with a half-filled band are insulators \cite{RevModPhys.40.677}. In the simplest model of the correlated system, introduced by Hubbard \cite{rspa.1963.0204}, the transition from the band metal to the Mott-Hubbard insulator ground state is govern by a ratio of the on-site Coulomb repulsion $U$ to the bandwidth $W$. In the case of one electron per atom, it takes place if $U > W$ \cite{rspa.1964.0190}. Then the quasiparticle band splits onto lower and upper Hubbard subbands, and the system becomes an insulator. The transition occurs in the same crystalline structure and can be triggered by different stimuli leading to a change in the band filling factor or the bandwidth. In many d-metal compounds, the pronounced effects of strong electron-phonon interaction (EPI) indicate the insufficiency of the purely electronic picture. The problem turns out to be very entangled since both Coulomb and electron-lattice interactions can cause such correlation effects as a renormalization of electron and phonon spectra, intraband and interband redistribution of the spectral weight, an increase in the effective mass, and a tendency towards localization of charge carriers.

Most of the theoretical studies of the mutual influence between electron-electron and electron-phonon interactions are focused on the Hubbard-Holstein model, which includes local Coulomb contribution and on-site modulation of the particle potential via lattice vibrations. Its phase diagram contains \cite{PhysRevB.70.125114,PhysRevB.96.205141,PhysRevB.99.045118,PhysRevLett.99.146404,PhysRevLett.125.167001} a metallic ground state when both interactions are small, Mott-Hubbard or charge-density wave insulating states, driven by Coulomb or electron-phonon interactions, respectively, and a variety of phases caused by competing orders. Here, we essentially consider an extension of this model, accounting for the multiband effects and off-site electron-phonon coupling, which modulates the kinetic energy of charge carriers. It has been shown earlier that multiorbital contributions significantly modify the phase diagram of the Hubbard-Holstein model \cite{PhysRevB.95.121112} at half-filling. For small Coulomb coupling and intermediate or strong electron-lattice one, the presence of nonequivalent bands leads to the occurrence of an orbital-selective insulating Peierls phase, which precedes the transition to the insulating state of charge density wave. In this phase, we can expect a non-Fermi-liquid behavior of charge carriers, similar to that is observed \cite{PhysRevLett.95.206401,PhysRevLett.102.126401} in the orbital-selective Mott phase \cite{Anisimov2002,PhysRevLett.92.216402} for Ising Hund’s coupling.

The nontrivial behavior of the system can also result from the Peierls or Su–Schrieffer–Heeger type contribution of the electron-phonon interaction. The modulation of orbital overlaps is characterized by the coupling parameter depending on both boson momentum and particle one and being off-diagonal in a real space representation. Such formulation of the problem goes beyond the applicability of the Gerlach-L{\"o}wen theorem, which rules out the non-analyticities in the ground state properties of the polaron system if coupling strength is constant or depends on boson momentum only~\cite{RevModPhys.63.63}. The revision of the concept of smoothly varying polaron properties starts from the demonstrations of the non-analyticity in the entanglement entropy of the polaron system \cite{PhysRevB.78.214301} and the sharp transition in the ground state energy of the single polaron \cite{PhysRevLett.105.266605}. Concerning the metal-insulator transition caused by the electron-lattice coupling in underdoped compound, the competition between its on-site and off-site contributions can determine the type of transformations, namely, amplitude and rate of the changes in the density of charge carrier states at the Fermi level~\cite{PhysRevB.101.235114}. Another critical consequence of the Peierls contribution is the formation of light bipolarons, which are stable against large values of the screened Coulomb repulsion~\cite{PhysRevLett.105.266605,PhysRevLett.121.247001}.

In this paper, we are interested in the general features of the insulator-metal transitions driven by doping in the particle-phonon coupled systems. We restrict our attention to the limit of strong electron correlations $U > W$ and assume the phonon field energy inherent for most compounds, $\omega_0 \ll t$. Our systematic analysis is based on the two-dimensional three-orbital pd model, which is relevant primarily for layered copper oxides. Nevertheless, it illustrates some evolution lows of transition behavior in correlated materials. It turns out that doping of a half-filled system can induce abrupt or smooth transitions with a variable amplitude of changes in the density of charge carrier states at the Fermi level. Moreover, different regimes of band structure transformations and corresponding types of transition behavior can be identified on the phase diagram of the system through the crossovers in polaron and bipolaron properties.

The total Hamiltonian reads $H = H^{el} + H^{ph} + H^{epi}$, where
\begin{eqnarray}
  \label{H_el}
H^{el}   &= &  \sum\limits_{{\bf{g}},\sigma } {\left( {{\varepsilon _d} - \mu } \right)n_{{\bf{g}},\sigma }^d}  + \sum\limits_{\bf{g}} {{U_d}n_{{\bf{g}},\sigma }^dn_{{\bf{g}}, - \sigma }^d}  + \nonumber \\
  && +   \sum\limits_{{\bf{g}},{\bf{r}},\sigma } {\left( {{\varepsilon _p} - \mu } \right)n_{{\bf{g}} + {\bf{r}},\sigma }^p}  + \sum\limits_{{\bf{g}},{\bf{r}}} {{U_p}n_{{\bf{g}} + {\bf{r}},\sigma }^pn_{{\bf{g}} + {\bf{r}}, - \sigma }^p}  + \nonumber \\
&& +  \sum\limits_{\left\langle {{\bf{g}},{\bf{g}}'} \right\rangle ,{\bf{r}},{\bf{r}}',\sigma } {{P_{{\bf{r}}{{\bf{r}}}'}}{t_{pp}}\left( {p_{{\bf{g}} + {\bf{r}},\sigma }^\dag {p_{{\bf{g}}' + {\bf{r}}',\sigma }} + {\rm{H.c.}}} \right)}  + \nonumber \\
&& +   \sum\limits_{{\bf{g}},{\bf{r}},\sigma } {{P_{\bf{r}}}{t_{pd}}\left( {d_{{\bf{g}},\sigma }^\dag {p_{{\bf{g}} + {\bf{r}},\sigma }} + {\rm{H.c.}}} \right)} + \nonumber \\
&& +  \sum\limits_{{\bf{g}},{\bf{r}},\sigma ,\sigma '} {{V_{pd}}n_{{\bf{g}} + {\bf{r}} ,\sigma }^pn_{{\bf{g}},\sigma '}^d}, \\
\label{H_ph}
{H^{ph}} &=& \sum\limits_{\bf{g}} {\hbar \omega _{0} \left( {f_{\bf{g}}^\dag {f_{\bf{g}}} + {\textstyle{1 \over 2}}} \right)}, \\
\label{H_epi}
{H^{epi}} &=& \sum\limits_{{\bf{g}},\sigma } {{M_d}\left( {f_{\bf{g}}^\dag  + {f_{\bf{g}}}} \right)d_{{\bf{g}}\sigma }^\dag {d_{{\bf{g}}\sigma }}}  + \nonumber \\
 &+& \sum\limits_{{\bf{g}},{\bf{r}},\sigma } {{M_{pd}}{P_{\bf{r}}}\left( {f_{\bf{g}}^\dag  + {f_{\bf{g}}}} \right)\left( {d_{{\bf{g}}\sigma }^\dag {p_{{\bf{g}} + {\bf{r}},\sigma }} + {\rm{H.c.}}} \right)}.
\end{eqnarray}
The Hamiltonian~(\ref{H_el}) describes the low-energy physics of the CuO-plane, which is a common structural unit in complex copper oxides with a partially filled 3d-orbital. 
Here operators $d_{\bf{g},\sigma}^\dag $ and $p_{{\bf{g}}+{\bf{r}},\sigma}^\dag $ create a hole with spin $\sigma$ on $d_{x^2-y^2}$-copper and $p_{x\left( y \right)}$-oxygen atomic orbitals at positions indicated by vectors ${\bf{g}}$ or ${\bf{g}}+{\bf{r}}$, respectively. Vector $\bf{r}$ goes over oxygen atom positions in the square unit cell at site $\bf{g}$. The values  ${\varepsilon _d}$ and  ${\varepsilon _p}$ are the local energies of the corresponding atomic orbitals,  $n_{{\bf{g}},\sigma }^d $ and $n_{{\bf{g}} + {\bf{r}},\sigma }^p $ are the hole number operators, $t_{pp}$ and $t_ {pd}$ are the hopping parameters, $U_p$, $U_d$, and $V_{pd}$ are Coulomb  repulsion parameters, $\mu $ is the chemical potential, and ${P_{\bf{r}}}$ and $ {P_{{\bf{r}}{{\bf{r}}}'}} $ are the phase factors. 
In the free phonon term~(\ref{H_ph}), we consider the bond-stretching optical vibrations with energy $\omega_0$. The Hamiltonian $H^{epi}$ defines charge carriers that are linearly coupled  to this dispersionless mode through the modulation of the copper on-site energy and the copper-oxygen hopping energy. These electron-phonon interactions are characterized by $M_d$ and $M_{pd}$ parameters of the on-site or charge density displacement and the off-site or transitive contributions, respectively. 
Throughout the paper we use the following set of parameters: ${\varepsilon _{d}} = 0$,  $\varepsilon _{p} = 1.5$,  ${t_{pp}} = 0.86$, $t_{pd} = 1.36$, ${U_d} = 9$,  ${U_p} = 4$, ${V_{pd}} = 1.5$, $W=2.15$, and $ \omega _0 = 0.090$ (all in eV). The dimensionless electron-phonon coupling constants are defined as ${\lambda _{on\left( {off} \right)}} = {{M_{d\left( {pd} \right)}^2} \mathord{\left/  {\vphantom {{M_{d\left( {pd} \right)}^2} {W\hbar {\omega _0}}}} \right. \kern-\nulldelimiterspace} {W\hbar {\omega _0}}}$.

To study the problem, we employ the polaronic version of the generalized tight-binding (pGTB) method. A detailed description of this type of cluster perturbation theory is provided elsewhere~\cite{PhysRevB.92.155143,pGTB2018,PhysRevB.101.235114}. In contrast to the Lanczos method, we exactly determine all eigenstates of the CuO$_4$ cluster Hamiltonian and then discard the highest excited states in a controlled manner, preserving the character of the spectrum. Truncating the Hilbert space of phonons, we control the convergence of the ground and the first excited states energies with 0,1 or 2 holes per site with an error of no more than 1~$\%$. Optimization of the local basis allows us to efficiently compute the band structure of the correlated system from weak to strong electron-phonon coupling. Here, we simulate the intercluster contribution in the generalized mean-field approximation, taking into account the interaction of charge carriers with spin fluctuations. Each time we change the concentration of doped holes $x$ or the parameters of the electron-phonon interaction $\lambda _{on}$ and $\lambda _{off}$, the band structure and the chemical potential of the system are recomputed.

\begin{figure}
\center
\includegraphics[width=\linewidth]{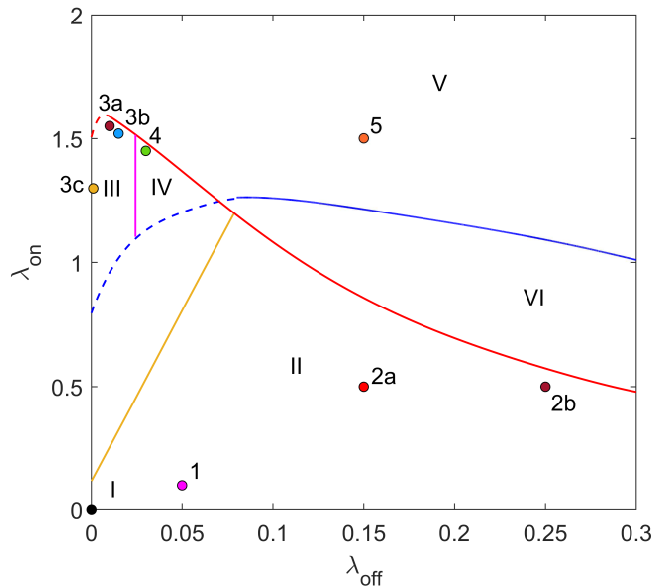}
\caption{\label{fig:f1}
Phase diagram of the system in the plane of on-site and off-site electron-phonon interaction parameters. The red and blue solid (dashed) curves trace the continuous (discontinuous) crossover in the properties of the polaron and bipolaron functions ${\left\langle {{n}_{1}^{d}} \right\rangle}_{0}$ (${\left\langle {{n}_{2}^{d}} \right\rangle}_{0}$), respectively. To the left (right) of the yellow curve, the on-site (off-site) contribution prevails. Between red and blue curves, the Roman numerals indicate the parameter regions with different types of the band structure transformations. The corresponding dependencies of the density of states on doping value are shown in the Figure~\ref{fig:f2}. }
\end{figure}

All results are collected below on the phase diagram in the plane of the on-site and off-site EPI parameters (Fig.~\ref{fig:f1}). To identify the patterns in the transition behavior, we compare them with polaron and bipolaron properties ${\left\langle {{n}_{1}^{d}} \right\rangle}_{0}$ and ${\left\langle {{n}_{2}^{d}} \right\rangle}_{0}$, characterizing the average number of charge carriers on the d-orbitals of copper in the single- and two-particle ground states of the unit cell cluster, respectively. Throughout the phase diagram, the total hole occupation of the copper orbitals increases with increasing on-site electron-lattice interaction. The corresponding evolution of the partial densities of states at the Fermi level is accompanied by continuous or discontinuous crossovers of the ${\left\langle {{n}_{1}^{d}} \right\rangle}_{0}$ and ${\left\langle {{n}_{2}^{d}} \right\rangle}_{0}$ functions with respect to the $\lambda_{on}$ parameter. We find that red and blue crossover curves in the Fig.~\ref{fig:f1} uniquely correlate with the type of the band structure transformation upon doping and, thus, determine the transition regimes.

\begin{figure}
\center
\includegraphics[width=\linewidth]{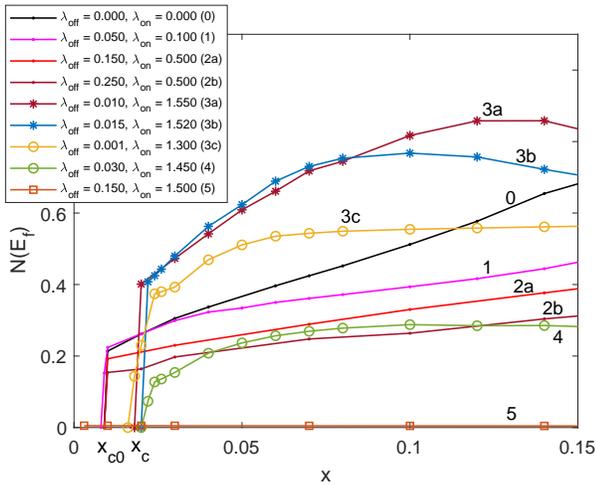}
\caption{\label{fig:f2}
Doping dependence of the density of states at the Fermi level for different types of the band structure transformations. Zero (curve 0) or weak (curve 1) electron-phonon interaction result in the sharp transitions at $ x=x_{c0} $. Intermediate electron-phonon interaction with dominating off-site contribution keeps the type of transition but reduces the final density of states $N(E_f)$ (curves 2a and 2b). The largest amplitude of changes in the density of states and sharp transition at $x_{c} > x_{c0}$ is observed with the prevailing on-site contribution (curves 3a and 3b). This transition is fast but smooth for purely on-site electron-lattice interaction (curve 3c). The formation of the flat band near the Fermi level can change the type of the transition and reduce its amplitude (curve 4). The strong interaction leads to the localization of charge carriers (curve 5).}
\end{figure}

First of all, we consider the system, neglecting the electron-phonon interaction (Fig.~\ref{fig:f2}, curve 0). At half-filling, narrow subbands of correlated d-electrons and a wide band of valence p-electrons of oxygen form the band structure of a charge-transfer insulator. The pd-hybridization leads to the mixing of the orbitals and the broadening of the correlated bands. Upon hole doping, we reveal the chemical potential in the gap near the bottom of the conduction band. It is located here up to some critical value of $x=x_{c0}$, where $x_{c0} \approx 0.8 \%$. For undoped system, such behavior reflects the absence of electron-hole symmetry. The latter guarantees the chemical potential is in the middle of the energy gap for a band insulator or a half-filled one-band Hubbard model \cite{PhysRevLett.43.1957,doi:10.1063/5.0035358}. The chemical potential confinement effect appears to be the result of the interband redistribution of the spectral weight caused by the Coulomb correlations. At sufficiently large value of the Coulomb interaction $U \ge 20$ eV or zero hybridization $t_{pd} = 0$, we find the  chemical potential near the top of the valence band. Then insulator-metal transition occurs for any non-zero value of the doped carriers, which is consistent with the paper \cite{PhysRevB.37.1597}. Anyway, the chemical potential gradually enters into the band upon doping, opening the Fermi surface with a maximum of the spectral weight at the $\left( {\pi }/{2}, {\pi }/{2} \right)$ point of the Brillouin zone. As a result, we observe a sharp transition from an antiferromagnetic Mott insulator to a metallic ground state, which is characterized by an abrupt increase in the density of states of charge carriers at the Fermi level $N(E_f)$. Weak electron-phonon interaction does not qualitatively affect the spectrum and behavior of the system (Fig.~\ref{fig:f2}, curve 1, Fig.~\ref{fig:f1}, part I). A slight smearing of the bands insignificantly decreases the spectral weight and density of states of quasiparticles emerging on the Fermi surface.

In contrast, strong electron-phonon interaction (Fig.~\ref{fig:f1}, part V) leads to the transition from itinerant to localized carriers at any fixed doping level and causes a rigid behavior of the system upon doping. In this regime, we do not observe a transition with an increase in the concentration of hole doped carriers up to $x=25-30\%$ (Fig.~\ref{fig:f2}, curve 5). A similar conclusion was previously obtained for the Hubbard-Holstein model using determinant Monte Carlo simulation~\cite{PhysRevB.96.205141}. In the phase diagram, the strong interaction part is located above both polaron and bipolaron crossover curves, where functions ${\left\langle {{n}_{1}^{d}} \right\rangle}_{0}$ and ${\left\langle {{n}_{2}^{d}} \right\rangle}_{0}$ reach their maximum values. Here, the distribution of the charge carrier density in the lattice among copper and oxygen orbitals takes the form of a checkerboard. Part VI of the phase diagram is characterized by moderately strong EPI effects. Between curves of singular crossovers in the properties of the functions ${\left\langle {{n}_{1}^{d}} \right\rangle}_{0}$ and ${\left\langle {{n}_{2}^{d}} \right\rangle}_{0}$, the system demonstrates transition to the state with a low density of electron carriers.

Intermediate electron-lattice coupling determines the doping transformations of the system between these two limits. 
To the right of the yellow line (Fig.~\ref{fig:f1}, part II), the off-site EPI contribution dominates. It is accompanied by a gradual increase of the band structure incoherence and subsequent modification of the transition driven by doping. Instead of sharp insulator-metal transition at weak electron-phonon coupling, we eventually observe a smooth transition to a semiconductor rather than a metallic state (Fig.~\ref{fig:f2}, curves 2a and 2b). To the left of the yellow line and between the crossover curves (Fig.~\ref{fig:f1}, part III and IV), the dominant on-site contribution of EPI causes qualitative changes in the band structure formation, which depend on the following circumstances. (i) The electron-phonon interaction develops a tendency towards localization of charge carriers on the d-orbitals of copper, the hybridization effects weaken, and the spectral weight of the quasiparticle excitations are redistributed. At half-filling, chemical potential gets stuck at Franck-Condon in-gap states with a low spectral weight near the bottom of the conduction band. Now the electron-phonon interaction restrains the transition up to some critical value of the doped carriers $x_{c} \gg x_{c0}$, where $ x_{c} \approx  2-3\%$ (Fig.~\ref{fig:f2}, curves 3-4). (ii) Simultaneously, a narrow flat band with a doping-dependent spectral weight begins to form around the $\left( \pi, \pi \right)$ point of the Brillouin zone~\cite{PhysRevB.101.235114}. For this parameter range, the phonon spectral function demonstrates the emergence of the novel polaronic states below the one-phonon continuum~\cite{PhysRevB.101.235114}. The flat band formation corresponds to the transitions between these excited polaronic states and ground bipolaronic state. This band is located just below the top of the valence band, but with an increase in the strength of the off-site electron-phonon coupling, its position gradually becomes higher. If the chemical potential passes through the gap under doping and first enters the valence band at the $\left( {\pi }/{2}, {\pi }/{2} \right)$ point, then a sharp insulator-metal transition occurs (curves 3a and 3b). According to the Gerlach-L{\"o}wen theorem, such transition is smooth, albeit very fast, for a purely local EPI contribution (curve 3c). Due to the multiparticle effects of the spectral weight redistribution, the amplitude of the changes in the density of states $N(E_f)$ now significantly exceeds that observed at low or zero EPI (Fig.~\ref{fig:f2}, curves 3a,3b and 3c versus curves 0-2). Otherwise, the chemical potential first falls into the flat band of impurity polaron states. This leads to a smooth increase in the density of states of quasiparticle excitations at the Fermi surface upon doping (Fig.~\ref{fig:f2}, curve 4).

\begin{figure}
\center
\includegraphics[width=\linewidth]{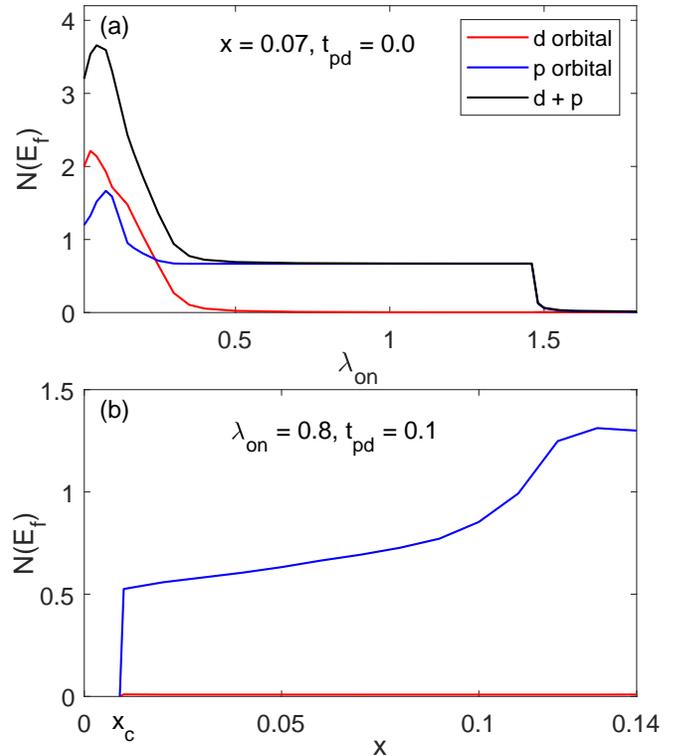}
\caption{\label{fig:f3}
Orbital-selective insulator-metal transition driven by (a) on-site electron-phonon interaction or (b) doping.}
\end{figure}

\begin{figure}
\center
\includegraphics[width=\linewidth]{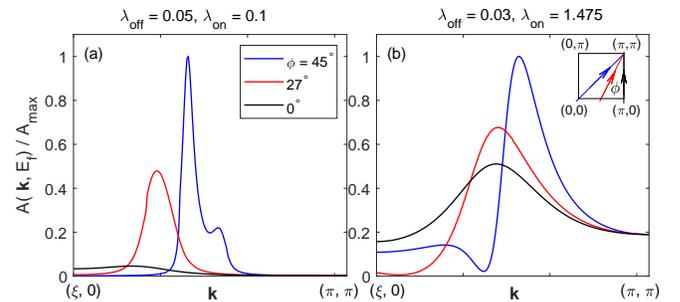}
\caption{\label{fig:f4}
Spectral weight mapping at the Fermi level as a function of the direction in the Brillouin zone for (a) weak and (b) intermediate EPI in the crossover regime (parts I and III of the phase diagram, respectively). The Fermi arc has no ``backside of the pocket'' if it results from the flat band formation (b).}
\end{figure}

It turns out the crossover regime (Fig.~\ref{fig:f1}, part III and IV) correlates with the mode of the orbital-selective insulator-metal transition observed at fixed doping level when any interorbital contributions are neglected (Fig.~\ref{fig:f3}a, $t_{pd}=0, V_{pd}=0, \lambda_{off}=0 $). Similar to the orbital-selective Mott transition~\cite{Anisimov2002}, an increase of the local electron-phonon interaction first localizes the charge carriers in the narrow d orbital at $ \lambda_{on} \ge 0.5 $, while the wide p band remains itinerant up to strong EPI, characterized by $ \lambda_{on} = 1.5 $. Any hybridization effects tends to suppress the orbital selective phase. With a fairly small contribution of the interorbital overlapping $t_{pd}$ we observe the orbital-selective insulator-metal transition driven by doping (Fig.~\ref{fig:f3}b, initial parameters of the band structure but $t_{pd} = 0.1$eV ), but there is no orbital-selective transition upon doping for the initial model parameter set. Nevertheless, the orbital-selective correlations persist and strongly influence the spectral weight redistribution between copper and oxygen orbitals at the dominating on-site electron-phonon contribution, especially in the crossover regime III and IV of the phase diagram.

Moreover, we find that the intermediate lattice contribution, which causes transitions in the crossover regime (Fig.~\ref{fig:f1}, part III and IV) at doping level $x=x_{c}$, leads to the pronounced pseudogap Fermi surface formation for $x \gg x_{c}$. Indeed, for the weak or zero electron-phonon interaction, spectral weight mapping in k space at the Fermi level demonstrates small hole pockets centered around $\left( {\pi }/{2}, {\pi }/{2} \right)$ point of the Brillouin zone (Fig.~\ref{fig:f4}a). It corresponds to the short-range antiferromagnetic spin liquid state of the system. However, for parts III and IV of the phase diagram (Fig.~\ref{fig:f1}), we observe \cite{PhysRevB.101.235114} short or elongated Fermi arcs centered around the point $\left( \pi, \pi \right)$ and having a maximum spectral weight at their centers (Fig.~\ref{fig:f4}b). The origin of the arches can differ and results from either the appearance of the flat band at the Fermi level or the spectral weight redistribution between the inner and outer sides of the pockets. The later effect is inherent for systems with strong electron correlations and is significantly enhanced here due to the orbital-selective correlations and flat band formation caused by electron-phonon interaction. 

Finally, we reveal different types of insulator-metal transitions driven by doping in the system with competing electron-electron and on-site and off-site electron-phonon interactions. In the limit of strong electron correlations and for the parameter range between the crossovers of the polaron and bipolaron properties, the intermediate electron-lattice coupling with dominating on-site contribution causes related effects such as orbital-selective behavior, pseudogap formation, emergence of the novel states below the one-phonon continuum in the phonon spectral function and sharp insulator-metal transitions with the largest amplitude of the change of the charge carriers density states at the Fermi level. This result directs the search for promising materials with exceptional characteristics of the dielectric-metal transition in correlated compounds and indicates the essential role of the on-site and off-site electron-lattice interactions in systems with strong Coulomb interaction. We expect qualitative changes in some parts of the phase diagram if there is no intersection of polaron and bipolaron crossovers. Such a case, as well as the effect of temperature, is the subject of future research.

\begin{acknowledgments}
The reported study was funded by Russian Foundation for Basic Research, Government of Krasnoyarsk Territory and Krasnoyarsk Regional Fund of Science according to the research project ''Studies of superexchange and electron-phonon interactions in correlated systems as a basis for searching for promising functional materials'' No. 20-42-240016.
\end{acknowledgments}

\bibliography{mybibfile}
\end{document}